# Molecular beam epitaxial growth of MoSe$_2$ on graphite, CaF$_2$ and graphene

*Suresh Vishwanath[1*], Xinyu Liu[2], Sergei Rouvimov[1], Patrick C. Mende[3], Angelica Azcatl[4], Stephen McDonnell[4], Robert M. Wallace[4], Randall M. Feenstra[3], Jacek K. Furdyna[2], Debdeep Jena[1] and Huili Grace Xing[1*]*

1. Electrical Engineering Department, University of Notre Dame, Notre Dame, IN, 46556, USA.

2. Physics Department, University of Notre Dame, Notre Dame, IN, 46556, USA.

3. Physics Department, Carnegie Mellon University, Pittsburgh, PA 15213, USA.

4. Department of Materials Science and Engineering, University of Texas at Dallas, Richardson, Texas 75083, USA

ABSTRACT: We report the structural and optical properties of molecular beam epitaxy (MBE) grown 2-dimensional (2D) material molybdenum diselenide (MoSe$_2$) on graphite, CaF$_2$ and epitaxial graphene. Extensive characterizations reveal that 2H- MoSe$_2$ grows by van-der-Waals epitaxy on all 3 substrates with a preferred crystallographic orientation and a Mo:Se ratio of 1:2.

Photoluminescence at room temperature (~1.56 eV) is observed in monolayer $MoSe_2$ on both $CaF_2$ and epitaxial graphene. The band edge absorption is very sharp, <60 meV over 3 decades. Overcoming the observed small grains by promoting mobility of Mo atoms would make MBE a powerful technique to achieve high quality 2D materials and heterostructures.

TEXT:

Layered materials have been at the center of attention since the discovery of graphene as they hold great promise for uncovering new physical phenomena and for creating new applications. Transition metal dichalcogenides (TMDs) are such materials systems possessing a wide range of energy bandgaps and band alignments. Some of the TMD materials have been shown to exhibit novel properties such as indirect to direct bandgap transition when their thickness is varied from few layers to monolayer [1,2], valley-polarized carriers [3,4,5], strain dependent bandgap variation [6,7,8] and more exotic properties like charge density waves [9] and superconductivity [10]. More recently, theoretical [11,12,13] and experimental [14] study of artificial stacking of these TMD materials is being pursued extensively to create heterostructures that are otherwise difficult to obtain in the conventional 3D epitaxy due to lattice constant mismatch. Most of these studies have been performed by exfoliating thin layers from natural or synthetic crystals [15,16] obtained using vapor phase transport technique (CVT) [17,18]. However, the manual stacking method makes the control of rotational orientation between the layered materials difficult. Typically, CVT grown and geological materials are unintentionally doped at rather high levels and the doping could vary spatially and correlated to surface defects potentially induced from the exfoliation process [18,19]. Lateral TMD heterostructures grown by chemical vapor deposition (CVD) have been recently reported but intermixing of the two materials is observed at the junction region [20]. Vertical heterostructure growth that simultaneously achieves spatial control as well as layer control while maintaining a large grain size is yet to be developed [21,22]. Molecular beam epitaxy (MBE) is widely used for a variety of material systems to obtain electronic grade materials with abrupt interfaces, thickness control and precise doping. Proof of concept MBE growth of TMD materials was demonstrated in early 1990s [23, 24, 25,26]. It was shown that 2D TMD thin films could be successfully grown on both 2D and 3D substrates by MBE and chemical beam epitaxy (CBE). More recently, the potential of MBE growth for TMD materials has been exhibited by the in-situ observation using ARPES [2] of direct to indirect transition of $MoSe_2$ with increase in layer thickness, as well as giant bandgap renormalization in monolayer $MoSe_2$ [27]. However, much work is yet needed to provide understanding of the resultant 2D crystal grain size, growth mechanisms and the effect of substrates. Such understanding is essential for preparing electronic grade materials that can enable high performance scalable devices. As a first step towards achieving electronic grade, high-quality 2D crystals, we present a growth study on one model material, $MoSe_2$, on highly oriented pyrolytic graphite (HOPG), $CaF_2$ (111) substrates with an inert surface terminated with fluorine, and epitaxial graphene on SiC.

HOPG, $CaF_2$ and epitaxial graphene on SiC are the three representative substrates used in this study. HOPG is a polycrystalline non-polar layered crystal with no out of plane bonds. $CaF_2$ is a polar 3D crystal with an inert surface termination. Electronic grade graphene (2D material) prepared on a single crystal substrate (SiC) is an ideal van-der-Waals substrate. Prior to loading into the MBE system, HOPG substrates (SPI Grade1) were cleaved using scotch tape to reveal a fresh surface for growth. The $CaF_2$ (Crystec) and epitaxial graphene on SiC substrates were cleaned sequentially in chloroform, acetone and methanol. All Substrates were first heated to 800 ºC in the growth chamber for 30 mins in vacuum (~ $5\times10^{-10}$ Torr) to allow desorption of weakly bound surface contaminants and then cooled to the growth temperature of 400 ºC. All growth temperatures in this manuscript are readings from the thermocouple behind the substrate holder. Once the growth temperature was stabilized, Mo and Se were deposited on the substrate simultaneously from the MBE sources in ultra high vacuum (UHV) conditions. Electron-beam evaporation was used for the Mo source, and a Knudsen cell was used for Se source. The Mo ion current was set to ~ 26 nA while the Se beam equivalent pressure (BEP) was maintained at ~ $6\times10^{-6}$ Torr. The growth rate determined by cross-sectional (cs) transmission electron microscopy (TEM) was ~0.3 monolayer per minute. The film thickness was varied from 0.6 - 9 monolayers (MLs) to investigate the film morphology, crystallinity and optical properties. After the MBE growth process, the excess selenium was removed in-situ by annealing the samples at 400 °C for ~5-10 minutes in the growth chamber with source shutters closed. The growth process was monitored in-situ by tracking the 10 keV reflection high-energy electron diffraction (RHEED) pattern and further analyzed ex-situ using low energy electron diffraction (LEED), low energy electron reflection (LEER), x-ray photoelectron spectroscopy (XPS), TEM, scanning electron microscopy (SEM), atomic force microscopy (AFM), Raman, photoluminescence (PL) and absorption spectroscopy. The samples for photoluminescence were annealed at 500 °C for 3 minutes and 600 °C for 7 minutes under Se flux before cooling down.

Figures 1a and 1b show the evolution of the RHEED pattern on HOPG. Before growth, the RHEED pattern of only HOPG is detected. After the growth of ~0.4 ML $MoSe_2$, the RHEED patterns corresponding to both $MoSe_2$ and HOPG are observed. The ratio of the in-plane lattice constants of $MoSe_2$ (0.3288 nm) and graphite (0.2461 nm) is 1.336. Since, RHEED pattern is an image in the reciprocal space, the inverse of the spacing ratio of the two sets of RHEED streaks gives a value of 1.333. This ratio is within 2% (within the error of the measurement) of the ratio of lattice constants of MoSe$_2$ and HOPG. This implies that the growth proceeds by van der Waals (vdW) epitaxy with no discernable strain. Crystallographically aligned growth of MoSe$_2$ to the surface orientation of the underlying substrate can be inferred from the fact that we observe RHEED patterns along the $\langle 11\bar{2}0 \rangle$ direction of MoSe$_2$ and HOPG. Similarly, crystallographically aligned growth is also observed for MoSe$_2$ grown on CaF$_2$ and epitaxial graphene. Figures 1c and 1d show that the RHEED streaks of MoSe$_2$ along $\langle 11\bar{2}0 \rangle$ appear at the same position as the $\langle 110 \rangle$ of CaF$_2$, as observed previously by Koma et al. [28]. This is further supported by the in-plane TEM diffraction of MoSe$_2$ on CaF$_2$ (Inset g2 of Fig. 3). We observe that the CaF$_2$ RHEED streaks vanish completely before the MoSe$_2$ RHEED streaks gradually appear. For an expected MoSe$_2$ growth of ~0.5 ML, RHEED patterns from neither material was observed. It is worthy to note that even though the CaF$_2$ RHEED streaks are sharp (Fig. 1c), the RHEED streaks from both 1.5 ML and 3 ML MoSe$_2$ on CaF$_2$ (Fig. 1d and Fig. S1c in the Supplementary Information (SI)) are blurry. This is in striking contrast to the sharper MoSe$_2$ RHEED streaks on HOPG for both 0.6 ML and 3.6 ML growth (Fig. 1b and Fig. S1a in the SI), suggesting a greater disorder in the as grown MoSe$_2$ on CaF$_2$ compared to MoSe$_2$ on HOPG. Removal of fluorine termination and increased reactivity on exposure to electron beam irradiation (i.e. electron stimulated desorption) has been confirmed by oxidation study by A.Koma et al. [28]. Therefore, care was taken to avoid the continuous exposure of the CaF$_2$ surface to the RHEED electron beam; only intermittent RHEED measurements were taken.

Fig. 1e-h shows a 2.4 ML MoSe$_2$ grown on epitaxial graphene on SiC sample characterized by low energy electron microscopy (LEEM). The LEEM image (Fig. 1e) shows a location at which both bare graphene and MoSe$_2$ on graphene coexist. Micro-LEED ($\mu$LEED) patterns from the graphene area and MoSe$_2$/graphene area indicate different crystal morphology, as shown in Fig. 1g and 1f, respectively. The epitaxial graphene is a single crystal with a perfect crystallographic orientation with the SiC underneath. Therefore, clear diffraction sets are seen; the MoSe$_2$/graphene area shows one set of hexagonal diffraction dots on top of a ring background. The observed ring background is consistent with a structure with a hexagonal lattice constant of 3.25 ± 0.02 Å, in agreement with the value of the MoSe2 in-plane lattice constant of 3.28 Å. This is also consistent with the RHEED pattern of MoSe$_2$ on HOPG (Fig. 1b). Therefore we conclude that the bright, continuous region seen in the LEEM image (Fig. 1e) is indeed covered by MoSe$_2$. With this identification, we can then explain the diffraction pattern of MoSe$_2$ on graphene. The size of the *µ*LEED aperture is 8 *µ*m and the diffraction pattern over this area results in a stronger set of hexagonal diffraction pattern overlaid on a faint ring. This indicates presence of many small grains, much smaller than the aperture size, most of which have a preferential orientation but some have random orientations. The preferential orientation in this case aligns with the orientation of the underlying graphene substrate. Additionally, Fig. 1f shows the low-energy electron reflectivity (LEER) spectra of monolayer graphene and bilayer graphene along with that of MoSe$_2$ on graphene. Here we see distinct differences between these spectra, specifically in the energy range of 0 - 6 eV. Graphene's LEER spectrum has a well-established evolution as a function of the number of monolayers present on the surface [29]. Concerning the MoSe$_2$ LEER spectrum, this is, as far as we are aware, the first such data presented on this material system. Based on previous work studying LEER of graphene with a first-principles method [30,31], we anticipate that the large interlayer spacing of MoSe$_2$ will result in a small hopping matrix element (which couples interlayer states localized between various layers). As a consequence, it is unlikely that LEEM will be capable of discriminating between different numbers of layers of TMDs in the same manner as for graphene. Nonetheless, LEEM has allowed us to confirm the presence of MoSe$_2$, as well as determine its preferential growth orientation as being aligned with the graphene underneath.

XPS was carried out on 2.4 ML MoSe$_2$ on HOPG to understand the stoichiometry of the as grown film. Figure 2a and 2b, show Se 3d$_{5/2}$ peak at 54.70 eV and Mo 3d$_{5/2}$ peak at 229.04 eV that are consistent with the formation of MoSe$_2$ [32]. When the Se:Mo ratio is calculated through deconvolution from the respective Mo and Se oxides, we obtain a ratio of 1.96, which is very close to the ideally expected stoichiometric value of 2. Also, no discernable signal corresponding to any excess elemental Se is observed in the XPS. When the take-off angle is varied we clearly observe that the oxide signal is greater at the surface than in the bulk. A hypothesis on the origin of this oxide is discussed in the SI. It is also noted that carbide formation is below the limit of detection, indicating that no covalent bonding of the MoSe$_2$ layer with HOPG is detected.

The TEM images of 9 ML MoSe$_2$ grown on HOPG and CaF$_2$ along with flakes exfoliated from bulk MoSe$_2$ are shown in Fig. 3. In the cs-TEM of MoSe$_2$ on HOPG (Fig. 3a), we observe a sharp interface between the conformal MoSe$_2$ film and HOPG. The interlayer spacing is calculated to be ~0.65 nm, which is very close to the reported value of 0.647 nm for bulk MoSe$_2$ [33]. Using FFT it is confirmed that the MoSe$_2$ crystal structure is indeed 2H and that the crystal plane perpendicular to the view direction is close to $(11\bar{2}0)$. In-plane TEM was performed by exfoliating MoSe$_2$ grown on HOPG to a TEM grid (Fig. 3b); small triangular domains of ~5 nm size stitched together are observed resulting in a near single crystal diffraction pattern locally (electron beam diameter ~150 nm). Formation of triangular grains during CVD growth of layered materials consisting of 2 different elements like h-BN, MoS$_2$, MoSe$_2$ etc. [34,35,36] has been previously observed. What is surprising here is the high degree of local rotational alignment because of van-der-Waals epitaxy in the MBE growth. Although such triangular features have recently been reported by other groups and studied using STM [37,38], the fact that these triangles are inherent in the as-grown material and not a Moiré pattern arising from interactions with the underlying substrate is evident from the diffraction pattern corresponding to HRTEM of MoSe$_2$ only (Fig. 3b, Inset). With increasing diameter of the electron beam, the diffraction spots get extended, gradually approaching the LEED pattern shown in Fig. 1h [27].

In the cs-TEM image of MoSe$_2$ grown on CaF$_2$ (Fig. 3f), two slightly misoriented grains stitched together are observed. The thin amorphous CaF$_2$ layer right below the MBE grown MoSe$_2$ is due to the loss of crystal structure during the TEM sample preparation, since CaF$_2$ is known to be very sensitive to radiation damage by electron or ion beam [28]. For thicker regions of the TEM sample (Inset f1 of Fig. 3f), we can see crystalline CaF$_2$ up to the interface with MoSe$_2$. Along the growth direction, the layers are well oriented in a 2H crystallographic form. However, in the growth plane a greater polycrystallinity is observed (Fig. 3g) as compared to MoSe$_2$ grown on HOPG (Fig. 3b). It is also observable in the in-plane TEM and the diffraction (Inset g1 of Fig. 3g) that majority of the MoSe$_2$ grains have a preferential orientation. But faint rings in the diffraction pattern are discernable, resulting from misoriented grains. The preferred orientation is aligned along the underlying CaF$_2$ crystal, as evident in Inset g2 of Fig. 3g. We can detect $\{220\}$ diffraction of CaF$_2$ (green dots) aligned with the $\{110\}$ or $\{11\bar{2}0\}$ diffraction of MoSe$_2$. This is consistent with the observed RHEED streaks of MoSe$_2$ along $\langle11\bar{2}0\rangle$ appear at the same position as the $\langle110\rangle$ of CaF$_2$ and with the relaxed growth as the lattice MBE grown MoSe$_2$ on CaF$_2$ is close to that of bulk MoSe$_2$. The fact that we do not observe all 6 spots corresponding to {220} of CaF$_2$ is due to tilt of the sample with respect to the zone axis. Finally, Fig. 3e shows the in-plane TEM of exfoliated MoSe$_2$ from a bulk sample that was imaged under the same conditions as the other in-plane TEM images. It proves that the features seen in MBE-grown material are intrinsic to the growth and not artifacts of the imaging.

The optical properties of the MBE MoSe$_2$ of varying thicknesses along with bulk MoSe$_2$ are shown in Fig. 4, including Raman, PL and absorption spectra. Similar to 2H-MoS$_2$, 2H-MoSe$_2$ belongs to the D$_{6h}$ group. Theoretical analysis predicts three Raman-active in-plane modes E$_{1g}$, E$^1_{2g}$, and E$^2_{2g}$, one active out-of-plane mode A$_{1g}$, and two inactive B$_{1u}$ and B$_{2g}$ modes [39]. In our experiment (Fig. 4a), few-layer and bulk MoSe$_2$ were analyzed using a 488 nm laser of 3 mW laser power. Strong E$^1_{2g}$ and A$_{1g}$ Raman peaks and weak E$_{1g}$ and B$_{2g}$ peaks are observed. Raman signal from bulk MoSe$_2$ has been observed to be much weaker as compared to few-layer MoSe$_2$ [34,40]. In case of bulk low laser power results in low signal to noise causing almost indiscernible E$^1_{2g}$ [40]. Hence, we used higher power and in Fig. 4a spectra were normalized with respect to A$_{1g}$ peak intensity. For comparison, We detect the A$_{1g}$ peak at ~244.2 cm$^{-1}$ and E$^1_{2g}$ peak at 286.1 cm$^{-1}$ and no B$_{2g}$ for bulk MoSe$_2$ in agreement with literature [34,41]. In a normal incident backscattering Raman setup on a basal plane as used in this work, the E$_{1g}$ mode is theoretically forbidden [42]. The peak observed at ~170 cm$^{-1}$ is assigned to the E$_{1g}$ peak of MoSe$_2$ as no other Raman peaks are expected theoretically at that value [43]. This E$_{1g}$ peak might arise due to a slight deviation from the laser beam normal incidence on the basal plane, a 2 photon process [44], or an appreciable crystallographic disorder. The inactive mode B$_{2g}$ has been reported to become Raman active in few layer 2H-MoSe$_2$ due to the breakdown of translation symmetry [40], which is perhaps the reason we observe it in the MBE grown materials. In 9 ML MoSe$_2$ on CaF$_2$ and HOPG we observe the $A_{1g}$ peak at ~242.1 cm$^{-1}$ (a red shift from the bulk) but the $E^1_{2g}$ peak are at 289.5 cm$^{-1}$ and 286.7 cm$^{-1}$ (a blue shift from the bulk), respectively. The likely explanation for the relative shifts observed in Raman peaks is a combination of various effects such as local heating [45], dielectric environment [46], breakdown of translation symmetry in these MBE MoSe$_2$ layers compared to bulk and small grain size in MBE grown material. Difference in local heating is due to the different thermal conductivity of different substrates. The broadening in both $A_{1g}$ and $E^1_{2g}$ peaks could be attributed to the small grain size of these MBE MoSe$_2$ films causing special localization of phonons [47]. The main Raman peak characteristics are summarized in Fig. 4b. The fact that there is no interlayer chemical interaction when MoSe$_2$ is grown on epitaxial graphene on SiC is confirmed by the Raman spectrum of epitaxial graphene before and after growth. Raman spectra from MoSe$_2$ and graphene are simultaneously observed, as shown in Fig. 4b. After the growth of MoSe$_2$, the 2D peak of graphene is shifted by about ~11 cm$^{-1}$. Shift in the graphene 2D peak (13 cm$^{-1}$) in a mechanically exfoliated MoS$_2$/exfoliated graphene/SiO$_2$ structure has been recently observed and attributed to in-plane compressive strain on graphene due to encapsulation of graphene by MoS$_2$ [48]. In our case, such strain or change in dielectric environment [46] could be used to explain the observed shift.

Photoluminescence (PL) from monolayer MoSe$_2$ grown by MBE on HOPG at 77K and epitaxial graphene on SiC at room temperature (RT) and 77K has been very recently reported [27]. Here we report the RT PL from monolayer MoSe$_2$ grown on CaF$_2$ and epitaxial graphene on SiC, shown in Fig. 4d. A PL peak at ~1.563 eV on graphene and ~1.565 eV on CaF$_2$ is measured, which is close to the reported value of ~1.57 eV at RT for exfoliated monolayer MoSe$_2$ on SiO$_2$ [49] and 1.55 eV at RT for MBE grown MoSe$_2$ on bilayer epitaxial graphene [27]. This is consistent with our earlier claim that the growth on CaF$_2$ also proceeds by van der Waals epitaxy and MoSe$_2$ does not chemically interact with the underlying substrate. It is worthy to note that 3 times higher laser power is necessary to obtain PL of about the same intensity from MoSe$_2$ on epitaxial graphene as compared to than on CaF$_2$, due to charge transfer from MoSe$_2$ to graphene. As shown below in Fig. 5, the nominal monolayer growth of MoSe$_2$ results in patched coverage since the 2$^{nd}$ layer starts to grow while the first layer has not fully coalesced. Therefore, the 1 ML MoSe$_2$ grown by MBE is not suitable for the large area absorption spectroscopy measurement, given a direct-indirect bandgap crossover is expected for 1 ML and 2 ML MoSe$_2$. The absorption coefficient (alpha) was measured on a 9 ML MoSe$_2$ on CaF$_2$ and plotted in Fig. 4. On the semi-log scale, a sharp band-edge with a 1000x increase in alpha over ~60 meV increase in the photon energy is observed, corresponding to a slope of about 20 meV/decade. A sharp density of states distribution near the band edge is critical for achieving sub-60 mV/dec steep slope transistor applications [13]. The bandgap of the 9 ML MoSe$_2$ is calculated to be 0.96 eV from a linear fit to alpha obtained from absorption spectroscopy plotted on a linear scale (inset Fig. 4e) [50], which is close to 1.08 eV reported in literature for the bulk MoSe$_2$ sample at room temperature, determined by liner fitting to square root of photocurrent measured [51]. This variation is probably due to the difference in measurement and fitting techniques.

In order to understand the surface morphology of the resulting MoSe$_2$ films, SEM (Fig. 5a and Fig. 5c) and AFM (Fig. S5b and Fig. S5d) characterizations were carried out on the 9 ML samples. These images show that high protrusions on HOPG and wrinkles on CaF$_2$ are formed on the surface. The cs-TEM image (Fig. 5b) shows one of these protrusions formed in MoSe$_2$ on HOPG, which is ~20 nm tall, much higher than the thickness of the grown MoSe$_2$. Through a close inspection of a series of such protrusions, it is found that when the surface step height variation in HOPG is on the order of several monolayers thick, the MoSe$_2$ domains on the two sides of the step interact to form these high aspect ratio protrusions. These are unlikely to be wrinkles due to the thermal expansion coefficient mismatch between MoSe$_2$ and HOPG since MoSe$_2$ has a positive thermal expansion coefficient: $a_a$ (basal-plane) of ~7.24x10$^{-6}$ °C$^{-1}$ and $a_c$ (out-of-plane) of ~12.93x10$^{-6}$ °C$^{-1}$ [52], while HOPG has a small and negative coefficient of linear expansion (~ -1x10$^{-6}$ °C$^{-1}$) in the basal plane below 400 °C [53]. Therefore, as the HOPG cools, it expands, whereas the MoSe$_2$ film shrinks. Hence, one would expect cracks rather than wrinkles/protrusions. In the case of CaF$_2$, its coefficient of linear thermal expansion is ~28 x10$^{-6}$ °C$^{-1}$ at the growth temperature (400 °C) and ~18 x10$^{-6}$ °C$^{-1}$ [54] at RT whereas for MoSe$_2$ it is less than half this value as noted earlier. Consequently, as the temperature is reduced, CaF$_2$ contracts much more than MoSe$_2$. Wrinkles are indeed observed (Fig. 5c and 5d) in the 9 ML MoSe$_2$ grown on CaF$_2$, similar to CVD growth of graphene [55]. Cs-TEM of the wrinkle (Fig. 5d) shows hollow space inside the wrinkle as expected.

To gain more understanding of the growth process, sub-monolayer (0.6 ML) growth of MoSe$_2$ on HOPG and epitaxial graphene was also carried out. A large density of nucleation was found, and both SEM and AFM images (Fig. 5e & 5f) clearly show that the 2$^{nd}$ layer grows before the 1$^{st}$ layer coalesces. The similar growth morphology was also very recently reported using scanning tunneling microscopy (STM) on MBE grown MoSe$_2$ on epitaxial graphene [27]. Both these growths can be explained by a low Mo adatom mobility due to the high Mo melting temperature and much lower temperature of the substrate in comparison. It is interesting to note that these flower shaped domains in Fig. 5e are formed by stitching of the much smaller triangular grains observed by TEM (Fig. 3b). In passing, it is also noted that Mo adatoms do move as we can observe greater nucleation along the step edges of HOPG (see Fig. S5). This observation could be used to design growth experiments to induce layer-controlled growth.

We have discussed the similarities and differences in MBE growth of MoSe$_2$ on HOPG and epitaxial graphene (van der Waals substrates) and CaF2 (quasi- van der Waals substrate due to inert surface fluorine termination). We observe that the growth occurs by van der Waals epitaxy in both cases and result in close to stoichiometric 2H oriented films. But the grains in the two cases are very different. Whether the underlying cause of this discrepancy is the quality of the substrate or something more fundamental is yet unclear. Raman features corresponding to MoSe$_2$ formation are observed. A shift of the 2D peak of graphene due to MoSe$_2$ is detected, implying an environmental dielectric interaction in spite of a lack of a detectable chemical interaction between the graphene substrate and MoSe$_2$. PL from monolayer MoSe$_2$ on CaF$_2$ at ~1.565 eV, on epitaxial graphene at ~1.563 eV and the bandgap of thick MoSe$_2$ of ~0.96 eV are measured, all at RT, very close to that from CVT grown MoSe$_2$. Finally, features resulting from growth of thick films on HOPG and CaF$_2$ have been investigated using SEM, AFM and cs-TEM. We believe this detailed study of the MBE grown TMD material, esp. using electron microscopy, in this paper would be a stepping-stone for design and benchmarking of MBE growth of 2D layered materials.

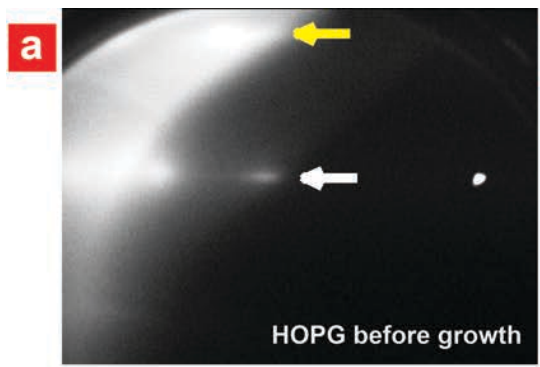
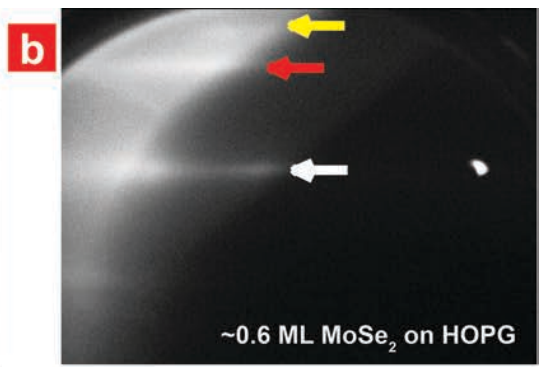
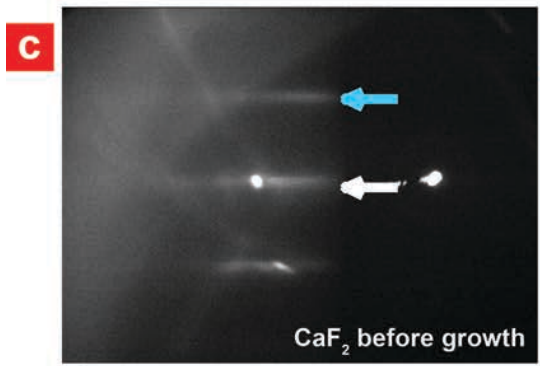
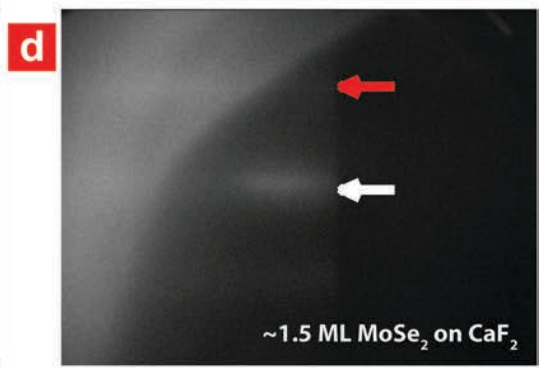
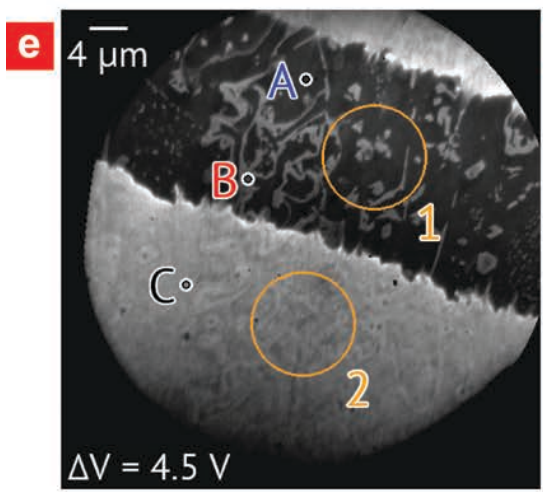
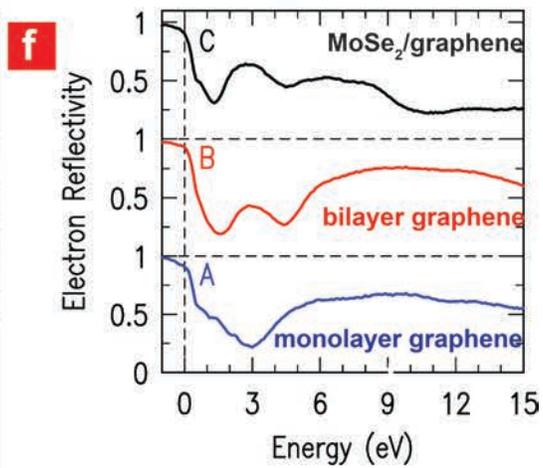
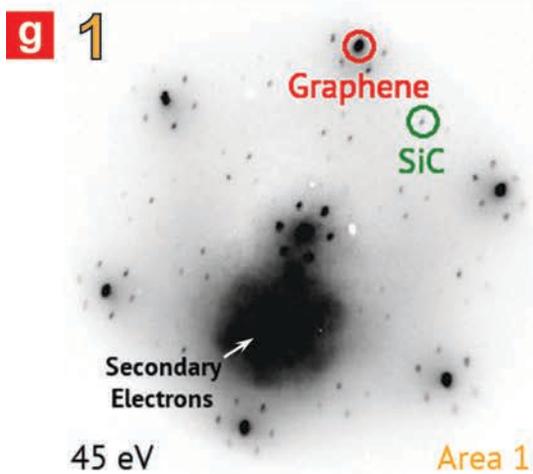
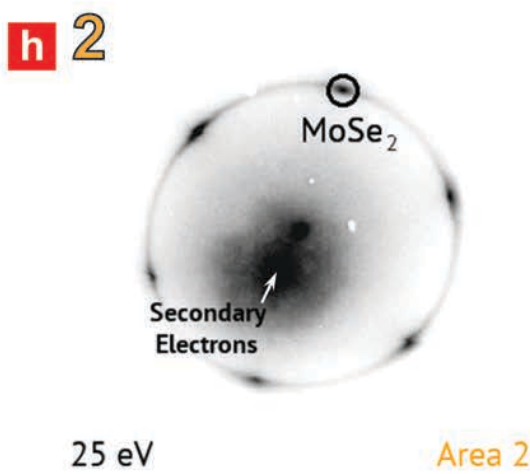

**Figure 1:** (a-d) RHEED images and (e-h) low electron energy analysis on 2.4 ML MoSe$_2$ on epitaxial graphene on SiC with part of the graphene substrate exposed. RHEED from (a) HOPG before growth, (b) ~0.6 ML MoSe$_2$ growth (Yellow arrow: HOPG and Red arrow: MoSe$_2$), (c) CaF$_2$ before growth, and (d) ~1.5 ML MoSe$_2$ growth on CaF$_2$. The RHEED behavior of MoSe$_2$ on epitaxial graphene is similar to that on HOPG, shown in Fig. S1. (e) LEEM image showing regions where LEER (f) and LEED (g and h) were performed. (f) LEER of monolayer graphene, bilayer graphene and 2.4 ML MoSe$_2$ on graphene. (g) LEED from graphene/SiC and (h) 2.4 ML MoSe$_2$ on graphene/SiC.

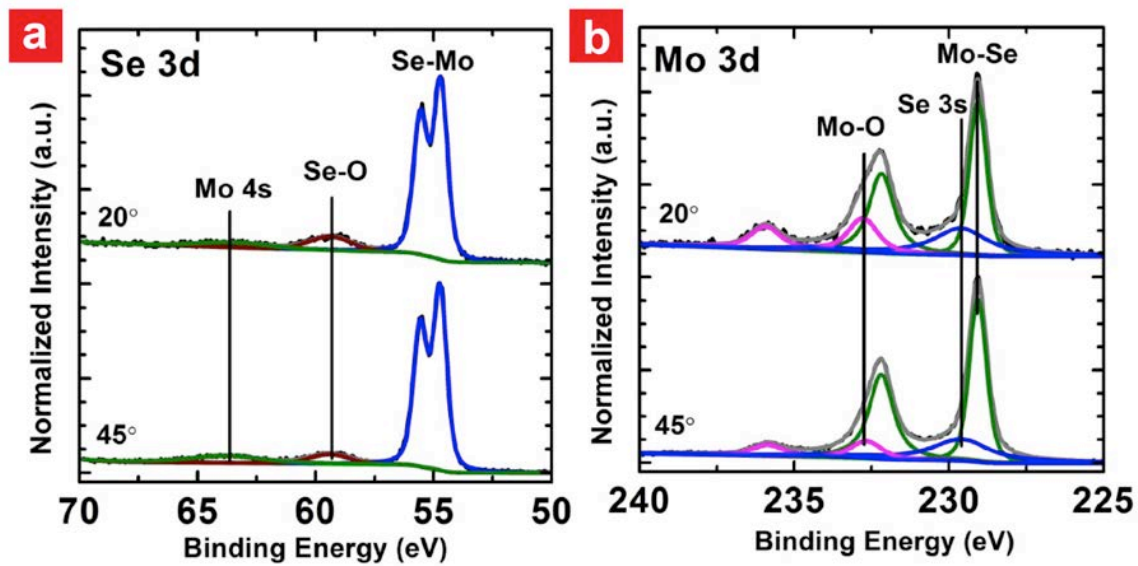

**Figure 2:** Angle resolved XPS spectra from 2.4 ML MoSe$_2$ grown on HOPG showing the (a) Se 3*d* and (b) Mo3*d* and Se 3*s* core levels, taken at 45° (bulk sensitive) and 20° (surface sensitive) take-off angles.

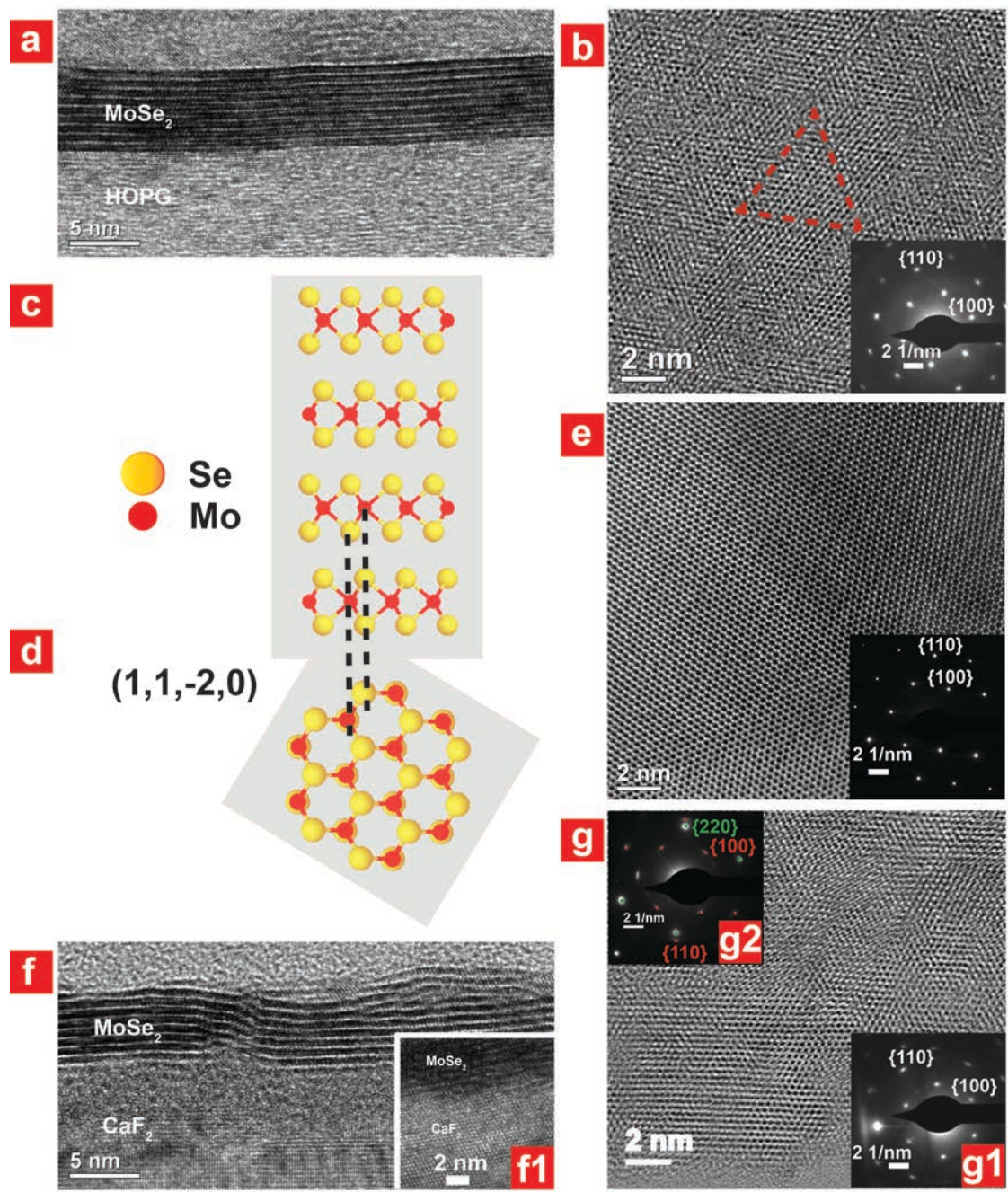

**Figure 3:** (a-b) 9 ML MoSe$_2$ on HOPG: (a) cs-TEM and (b) in-plane MoSe$_2$ TEM images with an inset showing the diffraction pattern from the same region. (c-d) Crystal model of 2H MoSe$_2$: (c) cross-section showing the $(11\bar{2}0)$ plane and (d) top view. (e) In-plane TEM along with the diffraction pattern of exfoliated flakes from bulk MoSe$_2$. (f-g) 9 ML MoSe$_2$ on CaF$_2$: (f) cs-TEM (inset f1 is from a thicker region of the TEM sample, showing crystalline CaF$_2$ and MoSe$_2$ interface.) and (g) in-plane MoSe$_2$ TEM images along with the diffraction pattern (inset g1). The other inset g2 shows diffraction patterns from a fragment of CaF$_2$ with MBE grown MoSe$_2$: the diffraction dots in green are for CaF$_2$ and the diffraction dots in red are for MoSe$_2$.

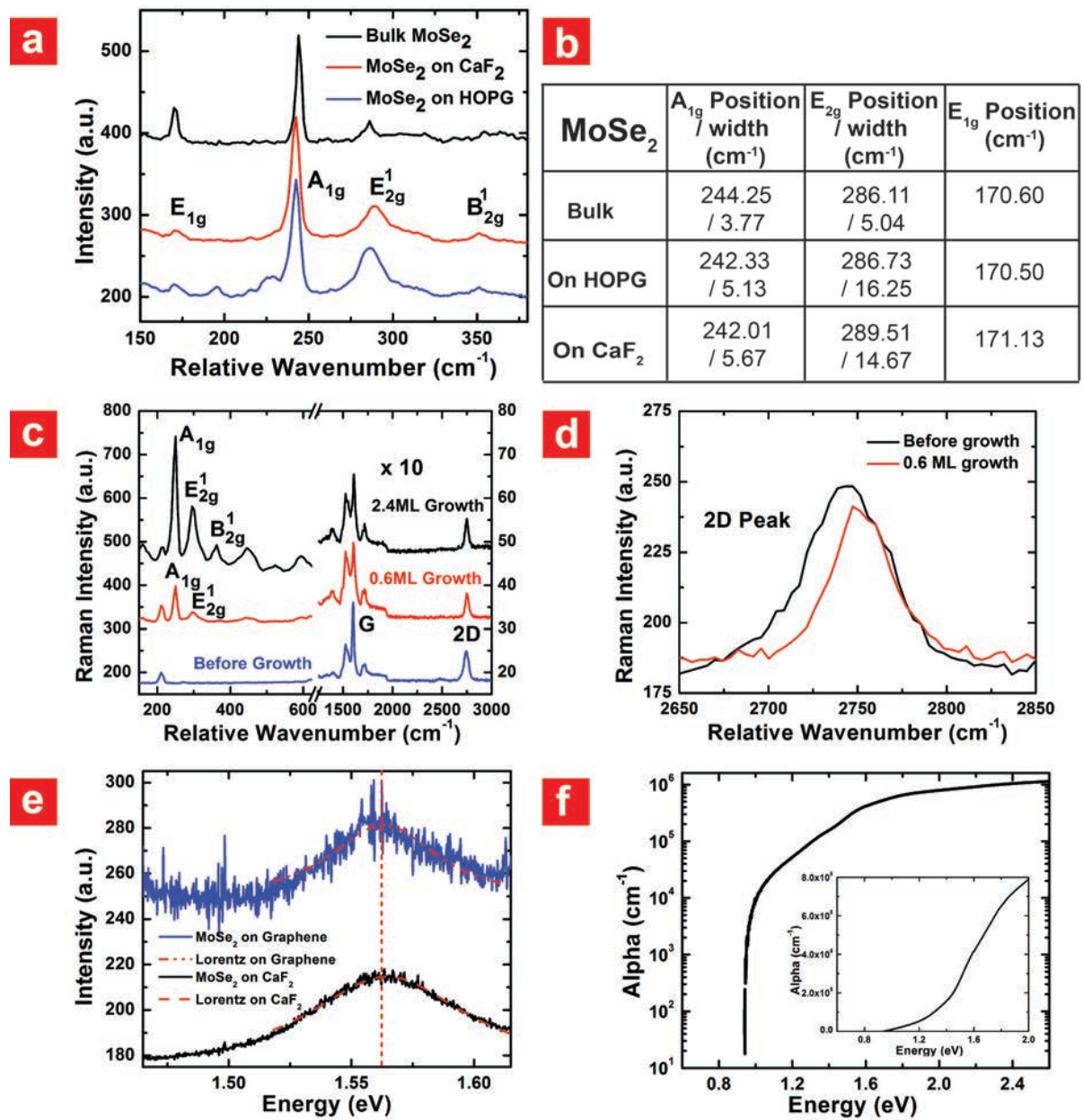

**Figure 4**: (a) Raman spectrum of 9 ML MoSe$_2$ grown on CaF$_2$ and HOPG compared to bulk MoSe$_2$. (b) lists the Raman peak positions obtained by Lorentzian fitting. (c) Evolution of Raman for MoSe$_2$ grown on epitaxial graphene/SiC, and (d) Raman shift in 2D peak of epitaxial graphene after growth of MoSe$_2$. (e) RT PL from ~1 monolayer MBE grown MoSe$_2$ on epitaxial graphene and CaF$_2$. (f) Semi-log plot of absorption coefficient measured on 9 ML MoSe$_2$ on CaF$_2$ (Inset: linear plot of the same data).

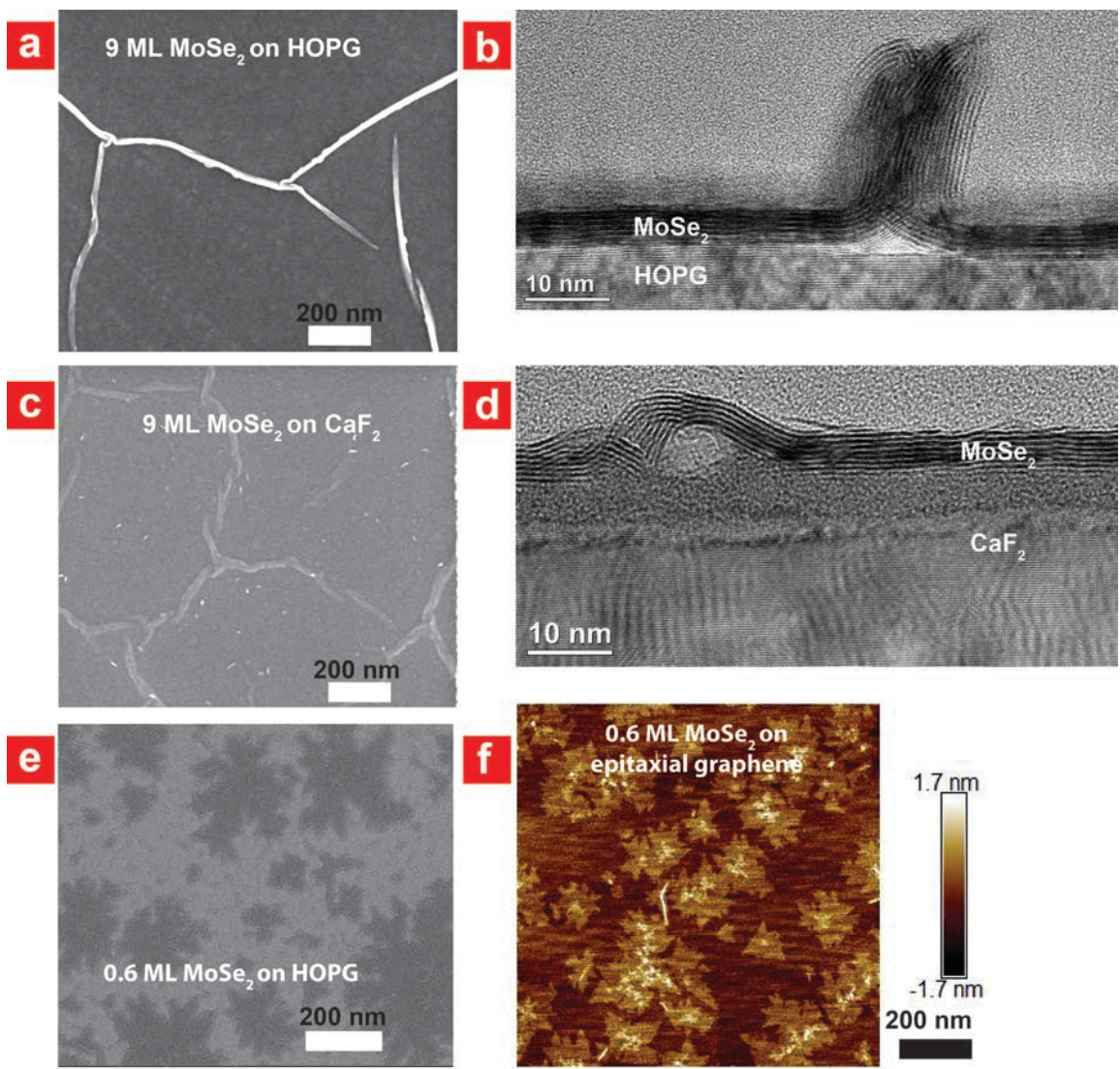

**Figure 5:** (a) SEM of the surface of 9ML MoSe$_2$ grown on HOPG (b) Cross-section of one of the protrusions of MoSe$_2$ on HOPG (c) SEM of the surface of 9 ML MoSe$_2$ grown on CaF$_2$ (d) Cross-section of one of the wrinkles of MoSe$_2$ on CaF$_2$ (e) SEM of the surface of 0.6 ML MoSe$_2$ grown on HOPG (e) AFM of 0.6 ML of MoSe$_2$ on epitaxial graphene on SiC.

ASSOCIATED CONTENT

**Supporting Information**. Additional information, methods, instrumentation and figures.

AUTHOR INFORMATION

**Corresponding Author**

* E-mails: svishwan@nd.edu; hxing@nd.edu

**Author Contributions**

The manuscript was written through contributions of all authors. All authors have given approval to the final version of the manuscript.

**Notes**

The authors declare no competing financial interest.

ACKNOWLEDGMENT

This work was supported in part by the Air Force Office of Scientific Research (AFOSR), the National Science Foundation (NSF), and the Center for Low Energy Systems Technology (LEAST), one of the six centers of STARnet, a Semiconductor Research Corporation program sponsored by MARCO and DARPA. We thank the ND Energy Materials Characterization Facility (MCF) for the use of the UV-Visible-Near IR Spectrometer and the ND Integrated

Imaging Facility for the use and understanding of Transmission Electron Microscope and Focus Ion Beam. The MCF is funded by the Sustainable Energy Initiative (SEI), which is part of the Center for Sustainable Energy at Notre Dame (ND Energy).

REFERENCES

(1) Mak, K. F.; Lee, C.; Hone, J.; Shan, J.; Heinz, T. F. *Phys. Rev. Lett.* **2010**, *105*, 136805.

(2) Zhang, Y.; Chang, T.-R.; Zhou, B.; Cui, Y.-T.; Yan, H.; Liu, Z.; Schmitt, F.; Lee, J.; Moore, R.; Chen, Y.; Lin, H.; Jeng, H.-T.; Mo, S.-K.; Hussain, Z.; Bansil, A.; Shen, Z.-X. *Nat. Nanotechnol.* **2014**, *9*, 111–115.

(3) Zeng, H.; Dai, J.; Yao, W.; Xiao, D.; Cui, X. *Nat. Nanotechnol.* **2012**, *7*, 490–493.

(4) Jones, A. M.; Yu, H.; Ghimire, N. J.; Wu, S.; Aivazian, G.; Ross, J. S.; Zhao, B.; Yan, J.; Mandrus, D. G.; Xiao, D.; Yao, W.; Xu, X. *Nat. Nanotechnol.* **2013**, *8*, 634–638.

(5) Wu, S.; Ross, J. S.; Liu, G.-B.; Aivazian, G.; Jones, A.; Fei, Z.; Zhu, W.; Xiao, D.; Yao, W.; Cobden, D.; Xu, X. *Nat. Phys.* **2013**, *9*, 149–153.

(6) Horzum, S.; Sahin, H.; Cahangirov, S.; Cudazzo, P.; Rubio, a.; Serin, T.; Peeters, F. *Phys. Rev. B* **2013**, *87*, 125415.

(7) Scalise, E.; Houssa, M.; Pourtois, G.; Afanas'ev, V.; Stesmans, A. *Nano Res.* **2011**, *5*, 43–48.

(8) Desai, S. B.; Seol, G.; Kang, J. S.; Fang, H.; Battaglia, C.; Kapadia, R.; Ager, J. W.; Guo, J.; Javey, A. *Nano Lett.* **2014**, *14*, 4592–4597.

(9) Staley, N.; Wu, J.; Eklund, P.; Liu, Y.; Li, L.; Xu, Z. *Phys. Rev. B* **2009**, *80*, 184505.

(10) Nagata, S.; Tsuyoshi, A.; Ebisu, S.; Ishihara, Y.; Tsutsumi, K. *J. Phys. Chem. Solids* **1993**, *54*, 895–899.

(11) He, J.; Hummer, K.; Franchini, C. *Phys. Rev. B* **2014**, *89*, 075409.

(12) Terrones, H.; López-Urías, F.; Terrones, M. *Sci. Rep.* **2013**, *3*, 1549.

(13) Li, M. (Oscar); Esseni, D.; Snider, G.; Jena, D.; Grace Xing, H. *J. Appl. Phys.* **2014**, *115*, 074508.

(14) Yu, W. J.; Li, Z.; Zhou, H.; Chen, Y.; Wang, Y.; Huang, Y.; Duan, X. *Nat. Mater.* **2013**, *12*, 246–252.

(15) Lee, C.-H.; Lee, G.-H.; van der Zande, A. M.; Chen, W.; Li, Y.; Han, M.; Cui, X.; Arefe, G.; Nuckolls, C.; Heinz, T. F.; Guo, J.; Hone, J.; Kim, P. *Nat. Nanotechnol.* **2014**, 8–13.

(16) Xiao, S.; Li, M.; Seabaugh, A.; Debdeep, J.; Xing, H. G. *Device Res. Conf.* **2014**, *72*, 169–170.

(17) Legma, J. B.; Vacquier, G.; Casalot, A. *J. Cryst. Growth* **1993**, *130*, 253–258.

(18) Nitsche, R.; Bolsterli, H. U.; Lichtenstriger, M. *J. Phys. Chem. Solids* **1961**, *21*, 199–205.

(19) Mcdonnell, S.; Addou, R.; Buie, C.; Wallace, R. M.; Hinkle, C. L. *ACS Nano* **2014**, *8*, 2880–2888.

(20) Huang, C.; Wu, S.; Sanchez, A. M.; Peters, J. J. P.; Beanland, R.; Ross, J. S.; Rivera, P.; Yao, W.; Cobden, D. H.; Xu, X. *Nat. Mater.* **2014**, 1–6. DOI: 10.1038/nmat4091

(21) Zhang, X.; Meng, F.; Christianson, J. R.; Arroyo-Torres, C.; Lukowski, M. a; Liang, D.; Schmidt, J. R.; Jin, S. *Nano Lett.* **2014**, *14*, 3047–3054.

(22) Gong, Y.; Lin, J.; Wang, X.; Shi, G.; Lei, S.; Lin, Z.; Zou, X.; Ye, G.; Vajtai, R.; Yakobson, B. I.; Terrones, H.; Terrones, M.; Tay, B. K.; Lou, J.; Pantelides, S. T.; Liu, Z.; Zhou, W.; Ajayan, P. M. *Nat. Mater.* **2014**, 1–8. DOI: 10.1038/nmat4091

(23) Ohuchi, F. S.; Shimada, T.; Parkinson, B. a.; Ueno, K.; Koma, a. *J. Cryst. Growth* **1991**, *111*, 1033–1037.

(24) Koma, A.; Yoshimura, K. *Surf. Sci.* **1986**, *174*, 556–560.

(25) Koma, A.; Ueno, K.; Saiki, K. *J. Cryst. Growth* **1991**, *111*, 1029–1032.

(26) Tiefenbacher, S.; Sehnert, H.; Pettenkofer, C.; Jaegermann, W. *Surf. Sci.* **1994**, *318*, L1161–L1164.

(27) Ugeda, M. M.; Bradley, A. J.; Shi, S.-F.; da Jornada, F. H.; Zhang, Y.; Qiu, D. Y.; Ruan, W.; Mo, S.-K.; Hussain, Z.; Shen, Z.-X.; Wang, F.; Louie, S. G.; Crommie, M. F. *Nat. Mater.* **2014**, 1–5. DOI: 10.1038/nmat4061

(28) Koma, A.; Saiki, K.; Sato, Y. *Appl. Surf. Sci.* **1989**, *41/42*, 451–456.

(29) Hibino, H.; Kageshima, H.; Maeda, F.; Nagase, M.; Kobayashi, Y.; Yamaguchi, H. *Phys. Rev. B* **2008**, *77*, 075413.

(30) Feenstra, R.; Srivastava, N.; Gao, Q.; Widom, M.; Diaconescu, B.; Ohta, T.; Kellogg, G.; Robinson, J.; Vlassiouk, I. *Phys. Rev. B* **2013**, *87*, 041406.

(31) Feenstra, R. M.; Widom, M. *Ultramicroscopy* **2013**, *130*, 101–108.

(32) Ohuchi, F. S.; Parkinson, B. a.; Ueno, K.; Koma, a. *J. Appl. Phys.* **1990**, *68*, 2168.

(33) James, P. B.; Lavik, M. T. *Acta Crystallogr.* **1963**, *16*, 1183–1183.

(34) Shaw, J.; Zhou, H.; Chen, Y.; Weiss, N.; Liu, Y. *Nano Res.* **2014**, *7*, 1–7.

(35) Helveg, S.; Lauritsen, J.; Laegsgaard, E.; Stensgaard, I.; Norskov, J.; Clausen, B.; Topsoe, H.; Besenbacher, F. *Phys. Rev. Lett.* **2000**, *84*, 951–954.

(36) Guo, N.; Wei, J.; Fan, L.; Jia, Y.; Liang, D.; Zhu, H.; Wang, K.; Wu, D. *Nanotechnology* **2012**, *23*, 415605.

(37) Liu, H.; Jiao, L.; Yang, F.; Cai, Y.; Wu, X.; Ho, W.; Gao, C.; Jia, J.; Wang, N.; Fan, H.; Yao, W.; Xie, M. *Phys. Rev. Lett.* **2014**, *113*, 066105.

(38) Murata, H.; Koma, A. *Phys. Rev. B* **1999**, *59*, 10327–10334.

(39) Verble, J. L.; Wieting, T. J. *Phys. Rev. Lett.* **1970**, *25*, 362–364.

(40) Tonndorf, P.; Schmidt, R.; Böttger, P.; Zhang, X.; Börner, J.; Liebig, A.; Albrecht, M.; Kloc, C.; Gordan, O.; Zahn, D. R. T.; Vasconcellos, S. M. D.; Bratschitsch, R. *Opt. Express* **2013**, *21*, 4908–4916.

(41) Sekine, T.; Izumi, M.; Nakashizu, T.; Uchinokura, K.; Matsuura, E. *J. Phys. Soc. JAPAN* **1980**, *49*, 1069–1077.

(42) Wieting, T. J.; Verble, J. L. *Phys. Rev. B* **1971**, *3*, 4286–4292.

(43) Terrones, H.; Del Corro, E.; Feng, S.; Poumirol, J. M.; Rhodes, D.; Smirnov, D.; Pradhan, N. R.; Lin, Z.; Nguyen, M. a T.; Elías, a L.; Mallouk, T. E.; Balicas, L.; Pimenta, M. a; Terrones, M. *Sci. Rep.* **2014**, *4*, 1–9. DOI:10.1038/srep04215

(44) Hajiyev, P.; Cong, C.; Qiu, C.; Yu, T. *Sci. Rep.* **2013**, *3*, 2593.

(45) Yan, R.; Simpson, J. R.; Bertolazzi, S.; Brivio, J.; Watson, M.; Wu, X.; Kis, A.; Luo, T.; Walker, A. R. H.; Xing, H. G. *ACS Nano* **2014**, *8*, 986–993.

(46) Yan, R.; Bertolazzi, S.; Brivio, J.; Fang, T.; Konar, A.; Birdwell, A. G. *arXiv:1211.4136v2* **2013**, 1–17.

(47) Gouadec, G.; Colomban, P. *Prog. Cryst. Growth Charact. Mater.* **2007**, *53*, 1–56.

(48) Zhou, K.; Withers, F.; Cao, Y.; Hu, S.; Yu, G.; Casiraghi, C. *ACS Nano* **2014**, *8*, 9914–9924.

(49) Ross, J. S.; Wu, S.; Yu, H.; Ghimire, N. J.; Jones, A. M.; Aivazian, G.; Yan, J.; Mandrus, D. G.; Xiao, D.; Yao, W.; Xu, X. *Nat. Commun.* **2013**, *4*, 1474.

(50) Mak, K. F.; Lee, C.; Hone, J.; Shan, J.; Heinz, T. F. *Phys. Rev. Lett.* **2010**, *105*, 136805.

(51) Hu, S. Y.; Lee, Y. C.; Shen, J. L.; Chen, K. W.; Tiong, K. K.; Huang, Y. S. *Solid State Commun.* **2006**, *139*, 176–180.

(52) El-mahalawy, S. H.; Evans, B. L. *J. Appl. Crystallogr.* **1976**, *9*, 403–406.

(53) Kelly, B. T. *Carbon* **1991**, *29*, 721–724.

(54) Ganesan, S.; Srinivasan, R. *Proc. Ind. Acad. Sci.* **1958**, *A25*, 139–153.

(55) Zhu, W.; Low, T.; Perebeinos, V.; Bol, A. A.; Zhu, Y.; Yan, H.; Terso, J. *Nano Lett.* **2012**, *12*, 3431–3436.

# Supplementary information of

# Molecular beam epitaxial growth of MoSe$_2$ on graphite, CaF$_2$ and graphene

*Suresh Vishwanath[1*], Xinyu Liu[2], Sergei Rouvimov[1], Patrick C. Mende[3], Angelica Azcatl[4], Stephen McDonnell[4], Robert M. Wallace[4], Randall M. Feenstra[3], Jacek K. Furdyna[2], Debdeep Jena[1] and Huili Grace Xing[1*]*

1. Electrical Engineering Department, University of Notre Dame, Notre Dame, IN, 46556, USA.

2. Physics Department, University of Notre Dame, Notre Dame, IN, 46556, USA.

3. Physics Department, Carnegie Mellon University, Pittsburgh, PA  15213-3890

4. Department of Materials Science and Engineering, University of Texas at Dallas, Richardson, Texas 75083, USA

*Emails: svishwan@nd.edu; hxing@nd.edu

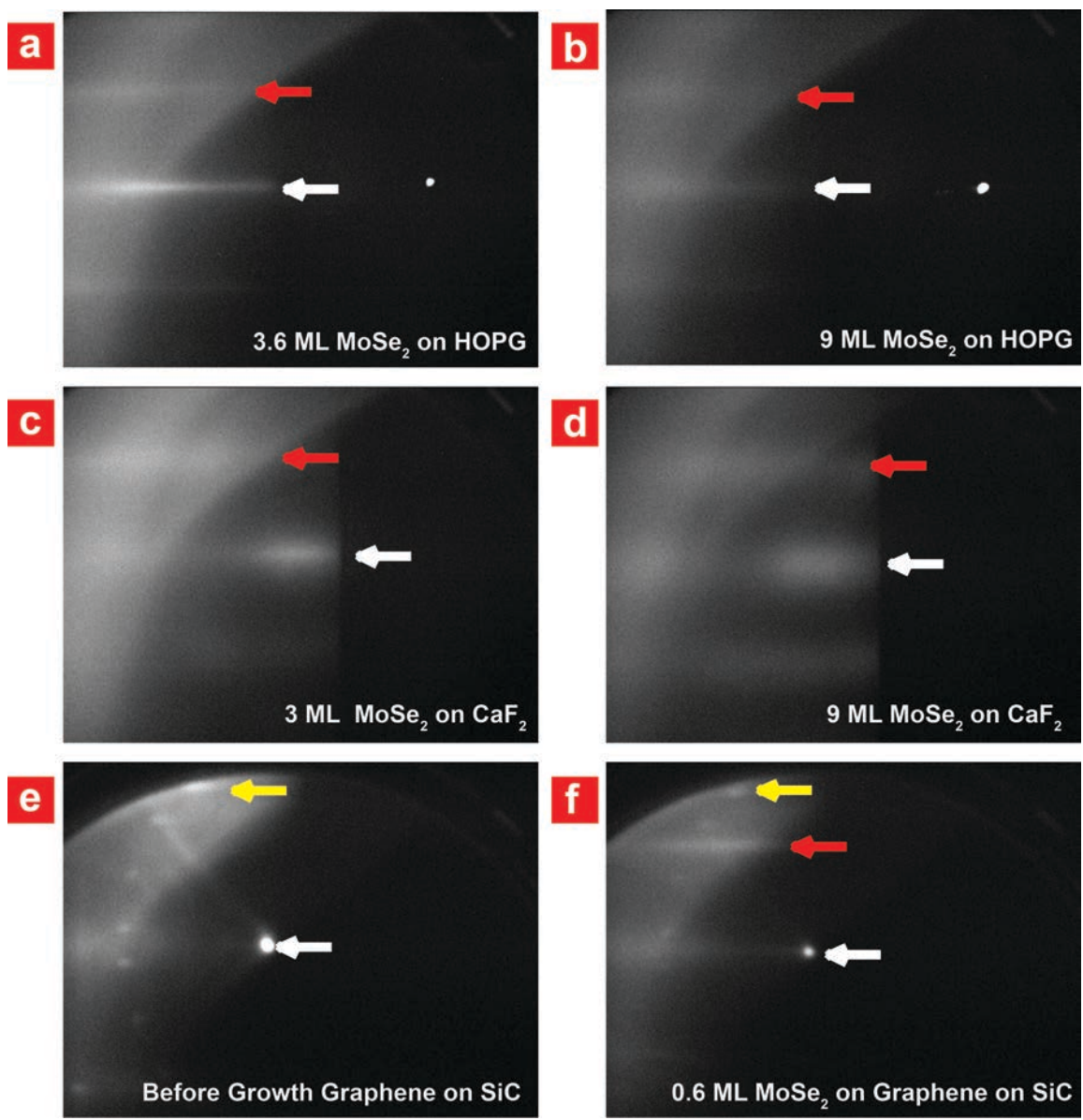

**Figure S1**: RHEED images from MoSe$_2$ growth on (a-b) HOPG, (c-d) CaF$_2$, and (e-f) epitaxial graphene on SiC (Red arrows: MoSe$_2$).

RHEED patterns are visible for 9 ML MoSe$_2$. However, RHEED intensity decreases appreciably with increasing number of layers in case of MoSe$_2$ growth on HOPG (Fig. S1a & b) while remains nearly constant for growth on CaF$_2$. Similar to the case of 0.6 ML growth of MoSe$_2$ on HOPG, RHEED on 0.6 ML MoSe$_2$ growth on epitaxial graphene, shows patterns from both MoSe$_2$ and graphene.

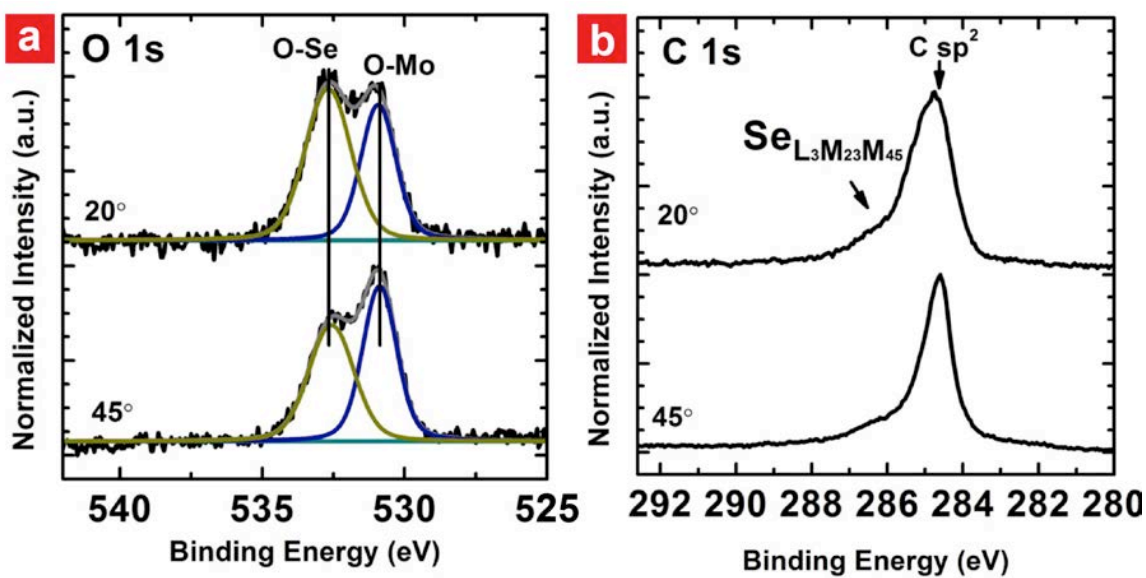

**Figure S2**: Angle dependent XPS showing the core levels of oxygen and carbon from MoSe$_2$ grown on HOPG.

We know from the AFM of sub-monolayer MoSe$_2$ on epitaxial graphene (Fig. 5f) and from previously reported STM on 0.8 ML MoSe$_2$ on epitaxial graphene [1] that the second layer grows before the completion of the first layer. Also, the domains formed by stitching of the small triangular gains observed in Fig. 3b are of the order of ~500 nm (Fig. 5e). Since, the XPS is done ex-situ on a 2.4 ML $MoSe_2$ on HOPG, we expect a large number of edge sites in $MoSe_2$. Higher reactivity of edges to oxygen [2] has been previously observed using XPS for bulk $MoSe_2$. Supported by the observation that with increasing incidence angle, a decrease in the percentage of oxygenated Mo and Se species from XPS decreases, which is in line with the greater population of edge sites at the surface as compared to the deeper layers, we believe that this oxidation is predominantly at these domain edge sites. It is also noted that there is no detectable carbide formation (~283 eV), indicating no detectable covalent bonding between the $MoSe_2$ and HOPG substrate.

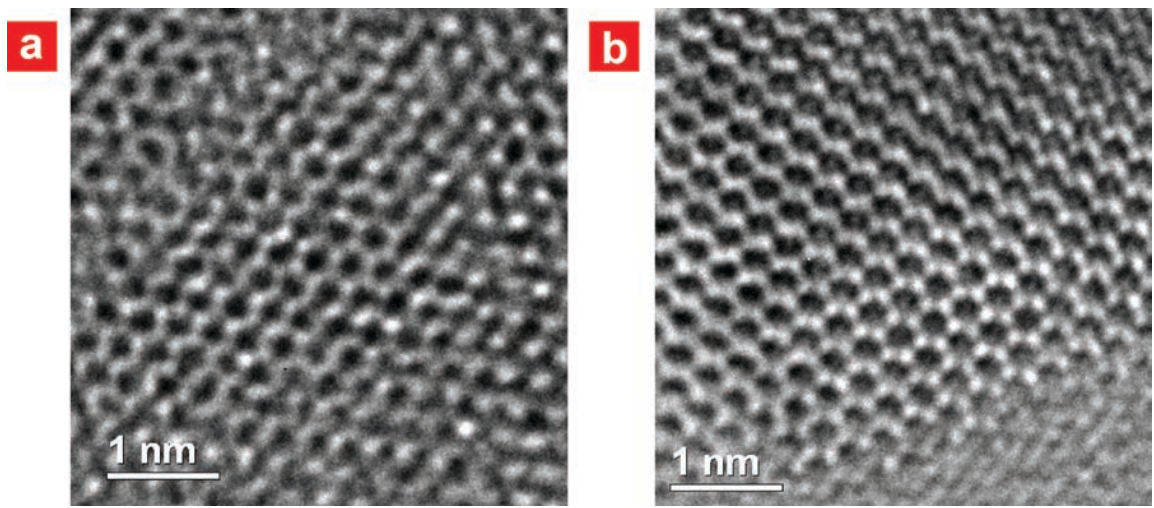

**Figure S3:** In-plane TEM images of MoSe$_2$ under similar imaging conditions: (a) grown on HOPG (b) exfoliated from bulk MoSe$_2$ We observe distorted crystal sites along the edges of the MBE MoSe$_2$ triangular domains. This is similar to the intrinsic structural defects observed in CVD MoS$_2$ [3] and is in striking contrast to the exfoliated MoSe$_2$ from the bulk.

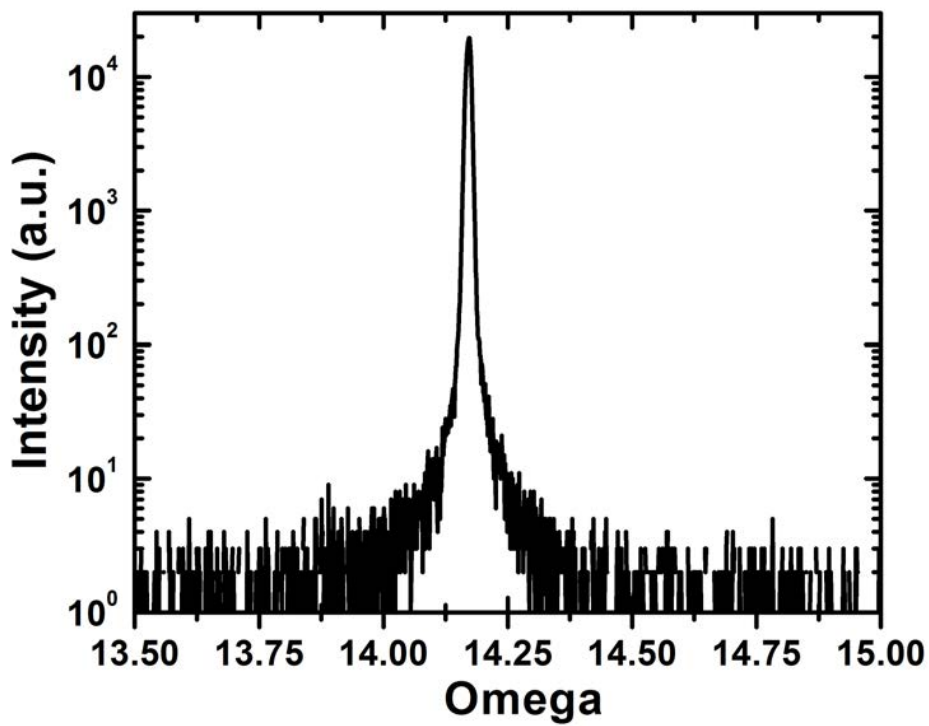

**Figure S4:** X-ray Diffraction rocking curve of $CaF_2$ substrate with a FWHM of 364.7 arcsec showing a well oriented (111) single crystal.

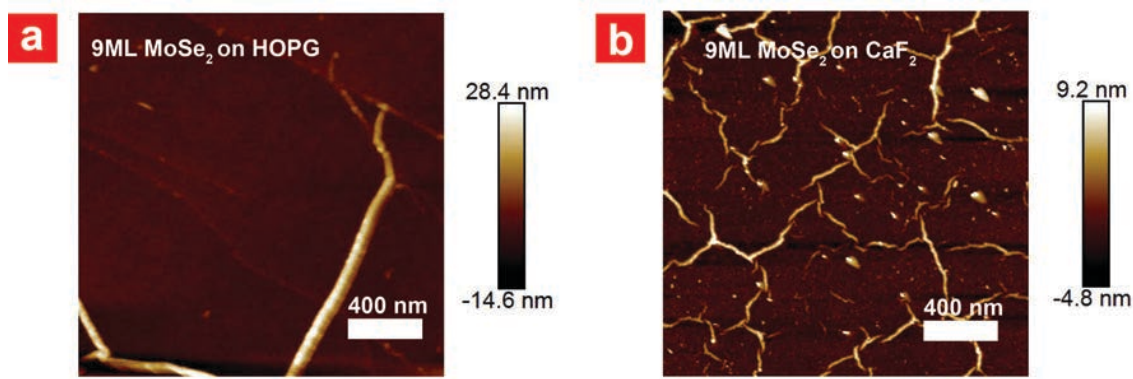

**Figure S5:** AFM of the surface of 9 ML $MoSe_2$ grown on (a) HOPG and (b) $CaF_2$.

The surface features on MoSe$_2$ on HOPG are much taller than those on MoSe$_2$ on CaF$_2$, supporting the hypothesis that the former are protrusions and the latter are wrinkles.

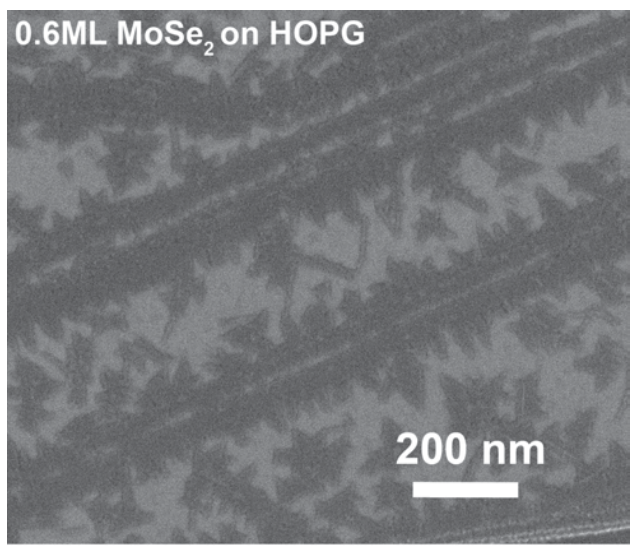

**Figure S6:** SEM of the surface of 0.6 ML MoSe$_2$ grown on HOPG, showing preferred nucleation along the steps on the substrate.

Methods and Instrumentation:

LEEM and LEED:

Low-energy electron microscopy (LEEM) and selected-area low-energy electron diffraction ($\mu$LEED) were performed with an Elmitec LEEM III. LEEM images are acquired in bright-field mode, wherein a contrast aperture is used to allow image formation only with those electrons that diffract into the (0,0) beam (i.e., only those electrons which diffract with zero lateral momentum). Reflectivity spectra are obtained by acquiring a sequence of images over a range of beam energies, and then extracting spectra at specific pixels in the image. $\mu$LEED is performed by collimating the electron beam through the introduction of an illumination aperture, restricting the beam size to approximately 8 $\mu$m.

Transmission Electron Microscopy:

The atomic structure analysis was carried out on FEI Titan 80-300 Transmission Electron Microscope operated at 300kV. Titan is capable of Transmission Electron Microscope (TEM) and Scanning Transmission Electron Microscope (STEM) modes with 2 A point-to-point resolution (below 1 A of information limit) and 0.136 nm resolution, respectively. The cross-sectional TEM samples have been prepared by Focus Ion Beam (FIB) technique at FEI Helios Dual Beam system. In-plane TEM sample from HOPG was prepared by careful mechanical exfoliation using scotch tape and from $CaF_2$ was prepared by sonication of the sample in methanol followed by drop casting on to the Cu TEM grid with holey carbon support film.

Raman and Photoluminescence:

Raman and Photoluminescence (PL) measurements were performed in the backscattering configuration using a WITec Alpha 300 system at room temperature. Following are the conditions used for different samples:

Raman of bulk, 9 ML MoSe$_2$ on HOPG and CaF$_2$: 100x objective, 1800 grooves/mm grating, 488 nm laser, 3mW, 2s/point, average over 10 accumulations.

PL of 1 ML MoSe$_2$ on CaF$_2$: 100x objective, 600 grooves/mm grating, 633 nm laser, 748μW, 10 s/point, average over 10 accumulations.

PL of 1 ML MoSe$_2$ on epitaxial graphene: 100x objective, 600 grooves/mm grating, 633 nm laser, 2.1mW, 10 s/point, average over 10 accumulations.

X-ray Photoelectron Spectroscopy:

XPS was carried out using a monochromated Al Kα source (hv = 1486.7 eV) and an Omicron Argus detector (MCD-128) operating with a pass energy of 15 eV. XPS spectra were acquired at a pass energy of 15 eV and take-off angle (defined with respect to the sample surface) of 45° and 20°. For XPS peak deconvolution, the spectral analysis software AAnalyzer was employed, where Voigt line shapes and an active Shirley background were used for peak fitting[4].

REFERENCES

(1) Ugeda, M. M.; Bradley, A. J.; Shi, S.-F.; da Jornada, F. H.; Zhang, Y.; Qiu, D. Y.; Ruan, W.; Mo, S.-K.; Hussain, Z.; Shen, Z.-X.; Wang, F.; Louie, S. G.; Crommie, M. F. *Nat. Mater.* **2014**, 1–5. DOI: 10.1038/nmat4061

(2) Chianelli, R. R.; Ruppert, A. F.; Behal, S. K.; Kear, B. H.; Wold, A.; Kershaw, R. *J. Catal.* **1985**, *92*, 56–63.

(3) Zhou, W.; Zou, X.; Najmaei, S.; Liu, Z.; Shi, Y.; Kong, J.; Lou, J.; Ajayan, P. M.; Yakobson, B. I.; Idrobo, J.-C. *Nano Lett.* **2013**, *13*, 2615–2622.

(4) Herrera-Gómez, a.; Hegedus, A.; Meissner, P. L. *Appl. Phys. Lett.* **2002**, *81*, 1014–1016.